\documentclass[preprint,amsmath,amssymb,aip,pop,longbibliography]{revtex4-2}
\usepackage[utf8]{inputenc}
\usepackage{amsmath}
\usepackage{amsfonts}
\usepackage{amssymb}
\usepackage{graphicx}
\usepackage{epstopdf}
\usepackage{graphicx}
\usepackage[subnum]{cases}
\usepackage{xcolor}

\newcommand{\lambdab}{{\bar\lambda}}
\def\fd#1,#2;{\frac{\delta #1}{\delta #2}}

\begin{document}

\title{Four-field Hamiltonian fluid closures of the one-dimensional Vlasov--Poisson equation}

\author{C. Chandre}
    \email{cristel.chandre@cnrs.fr}
    \affiliation{CNRS, Aix Marseille Univ, I2M, Marseille, France}

\author{B. A. Shadwick}
    \email{shadwick@unl.edu}
    \affiliation{University of Nebraska-Lincoln, Lincoln, NE, United States}

\begin{abstract}
We consider a reduced dynamics for the first four fluid moments of the one-dimensional Vlasov-Poisson equation, namely, the fluid density, fluid velocity, pressure and heat flux.  This dynamics
depends on an equation of state to close the system.  This equation of state (closure) connects the fifth order moment --related to the kurtosis in velocity of the Vlasov distribution-- with the first
four moments.  By solving the Jacobi identity, we derive an equation of state which ensures that the resulting reduced fluid model is Hamiltonian.  We show that this Hamiltonian closure allows
symmetric homogeneous equilibria of the reduced fluid model to be stable.
\end{abstract}

\maketitle

\section{Introduction}

In order to simulate the dynamics of a plasma, there is a variety of models which are used according to the type of question and the level of detail in the description of the plasma.  Most of these
models can be categorized as kinetic or fluid, whether the dynamical field variables are functions of the phase-space coordinates $(x,v)$ of the particles or just configuration space coordinates $x$.
Compared to kinetic models, fluid models have the significant advantage to be defined in a dimensionally reduced space, which makes them particularly desirable from a computational viewpoint.  The
central question is how to define these fluid models from a parent kinetic model.  There are plethora of methods to do this, some better suited than others depending on the specific problem at hand.
For instance, some reductions rely on an assumption on the shape of the distribution function,\cite{Gosse03,Jin03,Shadwick04,Fox09,Yuan11} or introduce suitably designed dissipative
terms.\cite{Hammett90,Passot04,Sarazin09}  Here we follow a different route by requiring that the reduced fluid model preserves an important dynamical property of the parent model, namely its
Hamiltonian structure.\cite{Perin14,Perin15,Perin15wb}  Rather than being an additional constraint on the reduction, we will see that this requirement provides a way to perform the reduction, and
precisely define the relevant closures leading to the definition of relevant Hamiltonian fluid model(s).  In order to illustrate this point we consider the one-dimensional Vlasov--Poisson equation.  This
equation describes the evolution of the distribution function $f(x,v,t)$ of charged particles (of charge $e$ and mass $m$) in an electric field $E(x,t)$:
$$
\frac{\partial f}{\partial t}=-v\frac{\partial f}{\partial x}-\frac{e\widetilde{E}}{m}\frac{\partial f}{\partial v},
$$
where $\widetilde{E}$ is the fluctuating part of the electric field $E$ whose dynamics is given by
$$
\frac{\partial E}{\partial t}=-4\pi \widetilde{j},
$$
and $j=e\int v f {\rm d}v$ is the current density. We assume periodic boundary conditions in $x$ with period $2L_x$, so that the fluctuating part is defined as
$$
\widetilde{E}=E-\frac{1}{2L_x}\int_{-L_x}^{L_x} E\; {\rm d}x.
$$
We consider a fluid description obtained by using the first four fluid moments of the distribution function, more precisely, the density $\rho(x,t)$, the fluid velocity $u(x,t)$, the pressure $P(x,t)$
and the heat flux $q(x,t)$ defined by
\begin{eqnarray*}
&& \rho = \int f {\rm d}v,\\
&& u = \rho^{-1}\int vf {\rm d}v ,\\
&& P=\int (v-u)^2 f {\rm d}v, \\
&& q = \frac{1}{2}\int (v-u)^3 f {\rm d}v.
\end{eqnarray*}
From the Vlasov-Poisson equation, we obtain the equations of motion for these moments and for the electric field:
\begin{subequations}
\label{eqn:motion}
\begin{eqnarray}
&&  {\partial_t} \rho=-{\partial_x} (\rho u),\\
&&  {\partial_t} u=-u {\partial_x} u-\frac{1}{\rho}\partial_x P +\frac{e\widetilde{E}}{m}\\
&&  {\partial_t}P= -u {\partial_x}P-3P {\partial_x}u -2\partial_x q,\\
&&  {\partial_t}q=-u {\partial_x}q-4q {\partial_x}u+\frac{3P}{2\rho}\partial_x P-\frac{1}{2}\partial_x R ,\\
&&  \partial_t E=-4\pi e \widetilde{\rho u},
\end{eqnarray}
\end{subequations}
where 
$$
R=\int (v-u)^4 f {\rm d}v,
$$
which is related to the kurtosis (in velocity) of the distribution function $f$.  Here and in what follows, $\partial_x$ and $\partial_v$ denote the partial derivatives of a function of $x$ and $v$
with respect to $x$ and $v$, respectively.  In order to close the set of equations of motion, we need an equation of state of the form
$$
R=R(\rho,u,P,q). 
$$
An example of closure is obtained by assuming a Gaussian distribution for $f$ (see Ref.~\onlinecite{Shadwick04}), 
$$
f(x,v,t)=\frac{\rho}{\sqrt{2\pi \sigma^2}}\,{\rm e}^{-(v-u)^2/(2\sigma^2)},
$$
which leads to $R=3P^2/\rho$, independent of $u$ and $q$.  One of the main problems of the Gaussian closure is that the resulting model breaks the original Hamiltonian structure of the parent model,
the Vlasov--Poisson equation.\cite{deGuillebon12} As a consequence, this closure introduces unphysical dissipation.

Based on the preservation of the Hamiltonian structure, another closure based on dimensional analysis was proposed in Ref.~\onlinecite{Perin15}, namely 
\begin{equation}
\label{eqn:Rda}
R=\frac{P^2}{\rho}+\frac{4q^2}{P}.
\end{equation}  
We notice that this closure depends explicitly on the asymmetries of the distribution function, measured by $q$, and is still independent of the fluid velocity $u$.  However this closure has a
fundamental drawback which is that homogeneous equilibria are all unstable.  In order to see this, we linearize the equations of motion around one of such equilibria with $q_0=0$, $u_0=0$ and $E_0=0$,
i.e., $\rho=\rho_0+\delta \rho$, $u=\delta u$, $P=P_0+\delta P$, $q=\delta q$ and $E=\delta E$.  The linearized equations of motion for $\delta {\bf X}=(\delta\rho, \delta u, \delta P,\delta q, \delta
E)$ in Fourier space, i.e., for
$$
\delta {\bf X}=\sum_{k=-\infty}^\infty \delta {\bf X}_k\, {\rm e}^{ikx},
$$
reduce to
\begin{equation}
\dot{\delta {\bf X}_k}=A\, \delta {\bf X}_k, 
\label{eqn:lin}
\end{equation}
where 
$$
A=\left( 
\begin{array}{ccccc}
0 & -i k \rho_0 & 0 & 0 & 0\\
0 & 0 & -i k \rho_0^{-1} & 0 & e/m \\
0 & -3i kP_0 & 0 & -2 i k & 0\\
i k\rho_0^{-2} P_0^2/2  & 0 & i k\rho_0^{-1}P_0/2 & 0 & 0\\
0 & -4\pi e\rho_0 & 0 & 0 & 0
\end{array} \right).
$$
The matrix $A$ does not have purely imaginary eigenvalues for 
$$
k^2<k_c^2=\frac{4\pi e^2\rho_0^2}{m P_0},
$$
from which we conclude that all equilibria with $q_0=0$ are unstable.  

Here we are looking for a closure which combines two important properties of the Vlasov--Poisson equation, namely, the stability of symmetric homogeneous equilibria, and its Hamiltonian structure. 

We do not assume any particular form for the distribution function.  Instead we solve the Jacobi identity in order to determine all possible $R(\rho,u,P,q)$ for which this identity is satisfied.  As a
result, we unveil a one-parameter family of Hamiltonian fluid closures.  We show that for these closures, the associated Poisson bracket has two Casimir invariants of the entropy type, i.e., two
observables $C$ of the form $C=\int {\rm d}x\ \rho\ \Gamma(\rho, P, q)$.  These Casimir invariants provide normal variables in which the closure in parametric form is found to be polynomial.  We then
examine numerically some properties of the resulting Hamiltonian model in two cases: plasma oscillations and the two-stream instability.

\section{Derivation of the four-field Hamiltonian closure}

The one-dimensional Vlasov--Poisson equation has a Hamiltonian structure~\cite{Morrison82} (see also Refs.~\onlinecite{Morrison98,Marsden02} for a review), i.e., the equations of motion can be recast
using a Hamiltonian and a Poisson bracket:
\begin{subequations}
\begin{align}
& \partial_t f=\{f,H\},\\
& \partial_t E=\{E,H\},
\end{align}
\end{subequations}
where
$$
H[f,E]=\int {\rm d}x {\rm d}v f \frac{m v^2}{2} +\int {\rm d}x \frac{E^2}{8\pi}.
$$
The Poisson bracket between two scalar functionals of $f$ and $E$ is given by
\begin{equation}
\label{eqn:VAbrack}
\{F,G\}=\frac{1}{m}\int f\left[\partial_x \fd F,f;\,\partial_v \fd G,f; -\partial_v \fd F,f;\,\partial_x \fd G,f;
-4\pi e \left(\widetilde{\fd F,E;}\,\partial_v \fd G,f;-\partial_v \fd F,f;\,\widetilde{\fd G,E;} \right) \right]{\rm d}x {\rm d}v,
\end{equation}
where $\fd F,f;\,$ and $\fd F,E;\,$ denote the functional derivatives of $F$ with respect to $f$ and $E$ respectively. In particular, this bracket satisfies the Jacobi identity, i.e.,
$$
\{F,\{G,H\}\}+\{H,\{F,G\}\}+\{G,\{H,F\}\}=0,
$$
for all observables $F$, $G$ and $H$. For simplicity of the notations and without loss of generality, we assume that $m=1$. 

{\em Remark:} Gauss's law is derived from a Casimir invariant of the bracket~\eqref{eqn:VAbrack}:
$$
C[f,E]=\partial_x E -4\pi e \int {\rm d}v f.
$$
Here we consider a neutral plasma, i.e., such that the value of this Casimir invariant is $-4\pi e$ which expresses the presence of a neutralizing background.  

Regardless of the truncation, Eq.~(\ref{eqn:motion}) can be recast in the following form (see Ref.~\onlinecite{Perin15} for more details):
$$
\partial_t {\bf X} = \{{\bf X},H\},
$$
where ${\bf X}=(\rho, u, S_2=P/\rho^3, S_3=2q/\rho^4, E)$, and 
$$
H[\rho, u, S_2,S_3, E]= \frac{1}{2}\int {\rm d}x \left[\rho u^2+\rho^3 S_2 +\frac{E^2}{4\pi}\right],
$$
and 
\begin{eqnarray}
\{F,G\}&=&\int {\rm d}x \left[\fd G,u;\, \partial_x \fd F,\rho;\, -\fd F,u;\, \partial_x \fd G,\rho;\, -4\pi e \left(\fd G,u;\, \widetilde{\fd F,E;\,} - \fd F,u;\, \widetilde{\fd G,E;\,}\right) 
\right. \nonumber \\[4pt]
&& \quad\left. -\frac{1}{\rho}\left(\fd G,u;\, \fd F,S_i;\, - \fd F,u;\, \fd G,S_i;\right)\, \partial_x S_i + \alpha_{ij}\,\frac1{\rho^2}\,\fd F,S_i;\,\fd G,S_j; +
\partial_x\left( \frac{1}{\rho}\,\fd F,S_i;\,\right)\beta_{ij}\frac{1}{\rho}\,\fd G,S_j;\right]. \label{eqn:PB}
\end{eqnarray}
The $2\times 2$ matrices $\alpha=\partial_x \gamma$ and $\beta$ are given by
$$
\gamma=\left(\begin{array}{cc}
2S_3 & 2S_4-3S_2^2\\ 3S_4-6S_2^2 & 3S_5-12S_2 S_3
\end{array} \right),
$$
and
$$
\beta= \left(\begin{array}{cc}
4S_3 & 5S_4-9S_2^2\\ 5S_4-9S_2^2 & 6S_5-24S_2 S_3
\end{array} \right),
$$
where $S_4=R/\rho^5$ and $S_5$ is an arbitrary function of $\rho$, $u$, $S_2$ and $S_3$.  As a consequence, since the bracket is antisymmetric, the models are all conserving energy regardless of the
closure $S_4=S_4(\rho,u,S_2,S_3)$ and $S_5=S_5(\rho,u,S_2,S_3)$.  We notice that $\beta=\gamma+\gamma^T$.  This allows us to
rewrite the Poisson bracket in a more antisymmetric way
\begin{eqnarray*}
\{F,G\}&=&\int {\rm d}x \left[\fd G,u;\, \partial_x \fd F,\rho;\, -\fd F,u;\, \partial_x \fd G,\rho;\, -4\pi e \left(\fd G,u;\, \widetilde{\fd F,E;\,}-\fd F,u;\, \widetilde{\fd G,E;\,}\right) \right. 
\nonumber \\[4pt]
&& \quad\left. {}- \frac1{\rho}\,\partial_x S_i\left(\fd G,u;\,\fd F,S_i; - \fd F,u;\, \fd G,S_i;\right) +
\frac1{\rho}\,\fd G,S_i;\,\gamma_{ij}\,\partial_x \left(\frac1{\rho}\,\fd F,S_j;\right) - \frac1{\rho}\,\fd F,S_i;\, \gamma_{ij}\,\partial_x\left(\frac1{\rho}\,\fd G,S_j;\right) \right].
\end{eqnarray*}
The Jacobi identity for the above bracket leads to the following constraints on the matrix $\gamma$:
\begin{subequations}
\label{eqn:jacgam}
\begin{align}
& (\gamma_{kn}+\gamma_{nk})\frac{\partial \gamma_{ij}}{\partial S_n}=(\gamma_{jn}+\gamma_{nj})\frac{\partial \gamma_{ik}}{\partial S_n},\\
& \frac{\partial \gamma_{in}}{\partial S_m}\frac{\partial \gamma_{jk}}{\partial S_n}=\frac{\partial \gamma_{jn}}{\partial S_m}\frac{\partial \gamma_{ik}}{\partial S_n},
\end{align}
\end{subequations}
for all $i$, $j$, $k$, $m$ (and repeated summation over $n$). 

\subsection{Explicit expression for the Hamiltonian closure}

In Ref.~\onlinecite{Perin15}, it was shown that in order for the bracket~(\ref{eqn:PB}) to be Hamiltonian, the closures $S_4$ and $S_5$ needs to be of the form $S_4=S_4(S_2,S_3)$ and $S_5=S_5(S_2,S_3)$, i.e., they do not depend on
$\rho$ and $u$.  The conditions~\eqref{eqn:jacgam} boil down to three constraints
\begin{eqnarray*}
&& 6S_5 =12S_2S_3+4S_3\frac{\partial S_4}{\partial S_2}+(5S_4-9S_2^2)\frac{\partial S_4}{\partial S_3},\\
&& \frac{\partial S_5}{\partial S_2}=4S_3 +\frac{\partial S_4}{\partial S_3}\left(\frac{\partial S_4}{\partial S_2}-3S_2 \right),\\
&& \frac{\partial S_5}{\partial S_3}=\frac{\partial S_4}{\partial S_2}+\left(\frac{\partial S_4}{\partial S_3}\right)^2.
\end{eqnarray*}
Equivalently, a necessary and sufficient condition is that the closure function $S_4$ satisfies the following two coupled nonlinear partial differential equations
\begin{subequations}
\label{eqn:Jac4}
\begin{align}
& 4S_3\frac{\partial^2 S_4}{\partial S_2^2}-(9S_2^2-5S_4)\frac{\partial^2 S_4}{\partial S_2\partial S_3}-\frac{\partial S_4}{\partial S_2}\frac{\partial S_4}{\partial S_3}-12S_3=0,\\
& 4S_3\frac{\partial^2 S_4}{\partial S_2\partial S_3}-(9S_2^2-5S_4)\frac{\partial^2 S_4}{\partial S_3^2}-\left(\frac{\partial S_4}{\partial S_3} \right)^2-2\frac{\partial S_4}{\partial S_2}+12S_2=0 .
\end{align}
\end{subequations}
From these equations, we readily check that the Gaussian closure $S_4=3S_2^2$ is not a solution of these equations, which means that the Gaussian closure is not Hamiltonian.  In addition, we check
that the solution given by Eq.~\eqref{eqn:Rda}, corresponding to the dimensional analysis of Ref.~\onlinecite{Perin15}, i.e., $S_4=S_2^2+S_3^2/S_2$, is the simplest solution.  However, this is not an
adequate solution since all homogeneous equilibria are always found to be unstable, as pointed out above.  To solve Eqs.~\eqref{eqn:Jac4}, we start by looking for solutions close to symmetric
distributions, i.e.,
$$
S_4(S_2,S_3)=f_0(S_2)+S_3^2 f_1(S_2)+O(S_3^4).
$$
We insert this expansion in Eqs.~(\ref{eqn:Jac4}) and consider their leading behavior near $S_3=0$. This lead to a set of two coupled ordinary differential equations
\begin{eqnarray*}
&& 2f_0''-(9S_2^2-5f_0)f_1'-f_1'f_0-6=0,\\
&& -(9S_2^2-5f_0)f_1-f_0'+6S_2=0.
\end{eqnarray*}
By combining these two equations, we obtain one single ordinary differential equation
$$
f_0''(9S_2^2-5f_0)+2f_0'^2-18S_2 f_0'+20f_0=0.
$$
Near $S_2=0$, we look for solutions of the type 
$$
f_0(S_2)=k S_2^\alpha. 
$$
A possible solution is obviously the one obtained using the dimensional analysis~\cite{Perin15}, i.e., $f_0(S_2)=S_2^2$.  In addition there is a less trivial family of solutions for $\alpha=5/3$.
More generally, we look at solutions which can be expanded in Puiseux series
$$
f_0(S_2)=\sum_{n=5}^\infty a_n S_2^{n/3}.
$$
We show that the only possible solutions are $f_0(S_2)=S_2^2$ and 
$$
f_0(S_2)=k S_2^{5/3},
$$
for any value of $k$.  For practical purposes, we define $\kappa=5k/9$.  We notice that contrary to the solution provided by dimensional analysis, the second solution comes as a family parameterized
by $\kappa$.  The interesting feature is that this family extends to a Hamiltonian closure for arbitrary large values of $S_3$.  Indeed we are looking for a solution which can be expanded as
\begin{equation}
\label{eqn:S4nc}
S_4(S_2,S_3)=\sum_{n=0}^\infty f_n(S_2)S_3^{2n}.
\end{equation}
Inserting this ansatz in Eq.~(\ref{eqn:Jac4}) leads to a recurrence relation for the coefficients $f_n(S_2)$: 
\begin{subequations}
\label{eqn:diff2}
\begin{align}
& f_0(S_2)=\frac{9\kappa}{5} S_2^{5/3},\\
& f_1(S_2)= \frac{\kappa  -2S_2^{1/3}}{3S_2(\kappa  -S_2^{1/3})},\\
& f_{n+1}=-\frac{1}{9(n+1)(2n+1)S_2^{5/3}(\kappa -S_2^{1/3})}\Biggl[(4n-1)f_{n}' \nonumber\\[4pt]
& \qquad \qquad \qquad \qquad +\sum_{m=1}^n m(12m-7-2n)f_{m}f_{n+1-m} \Biggr], 
\end{align}
\end{subequations}
and an addition constraint where $f_n$ has to satisfy 
\begin{eqnarray}
&& S_2^{5/3} (\kappa -S_2^{1/3})(n+1)f'_{n+1}-(n+1)\frac{\kappa}{3} S_2^{2/3}f_{n+1}+\frac{2}{9} f''_n\nonumber \\
&& \qquad +
\frac{1}{9}\sum_{m=1}^n (6m-n-1) f'_m f_{n+1-m}=0, \label{eqn:cond2}
\end{eqnarray}
for all $n\geq 1$. 
The first few terms are given by
\begin{eqnarray*}
&& f_2(S_2)= \frac{1}{3^4 S_2^{11/3}(\kappa  -S_2^{1/3})},\\[4pt]
&& f_3(S_2)= 2\frac{5\kappa -3S_2^{1/3}}{3^8 S_2^{19/3}(\kappa -S_2^{1/3})^{3}},\\[4pt]
&& f_4(S_2)= \frac{48\kappa^2-61\kappa S_2^{1/3}+18 S_2^{2/3}}{3^{11} S_2^9(\kappa-S_2^{1/3})^{5}}.
\end{eqnarray*}
The expression of other terms of the series expansion of $S_4$ can be obtained using a MATLAB~\cite{matlab} code available at Ref.~\onlinecite{github}.  We are not able to prove directly that for all
$n$, the $f_n$s obtained by the recursion relation~\eqref{eqn:diff2} satisfy Eq.~\eqref{eqn:cond2}.  However, we have checked that for $n$ below 25, these conditions are satisfied using symbolic
computations available from the MATLAB~\cite{matlab} code.  Beyond this value of 25, the symbolic computations are too complex to allow simplifications in a reasonable amount of time.  By truncating
the series~\eqref{eqn:S4nc}, i.e., by considering
$$
S_4(S_2,S_3)=\sum_{n=0}^{n_{\rm max}} f_n(S_2)S_3^{2n},
$$
we have found that the Jacobi identity is satisfied up to orders $S_3^{2 n_{\rm max}}$ for the values of $n_{\rm max}$ we have tested.  This led us to conjecture that the limit $n_{\rm max}\to \infty$
corresponds to a Hamiltonian closure.  We notice that the closure is singular at
$$
S_2^{(c)}= \kappa ^3,
$$
so this explicit closure $S_4=S_4(S_2,S_3)$ is valid only in the range $S_2\in [0,S_2^{(c)}[$. 

\begin{figure}
\begin{center}
\includegraphics[width=0.7\linewidth]{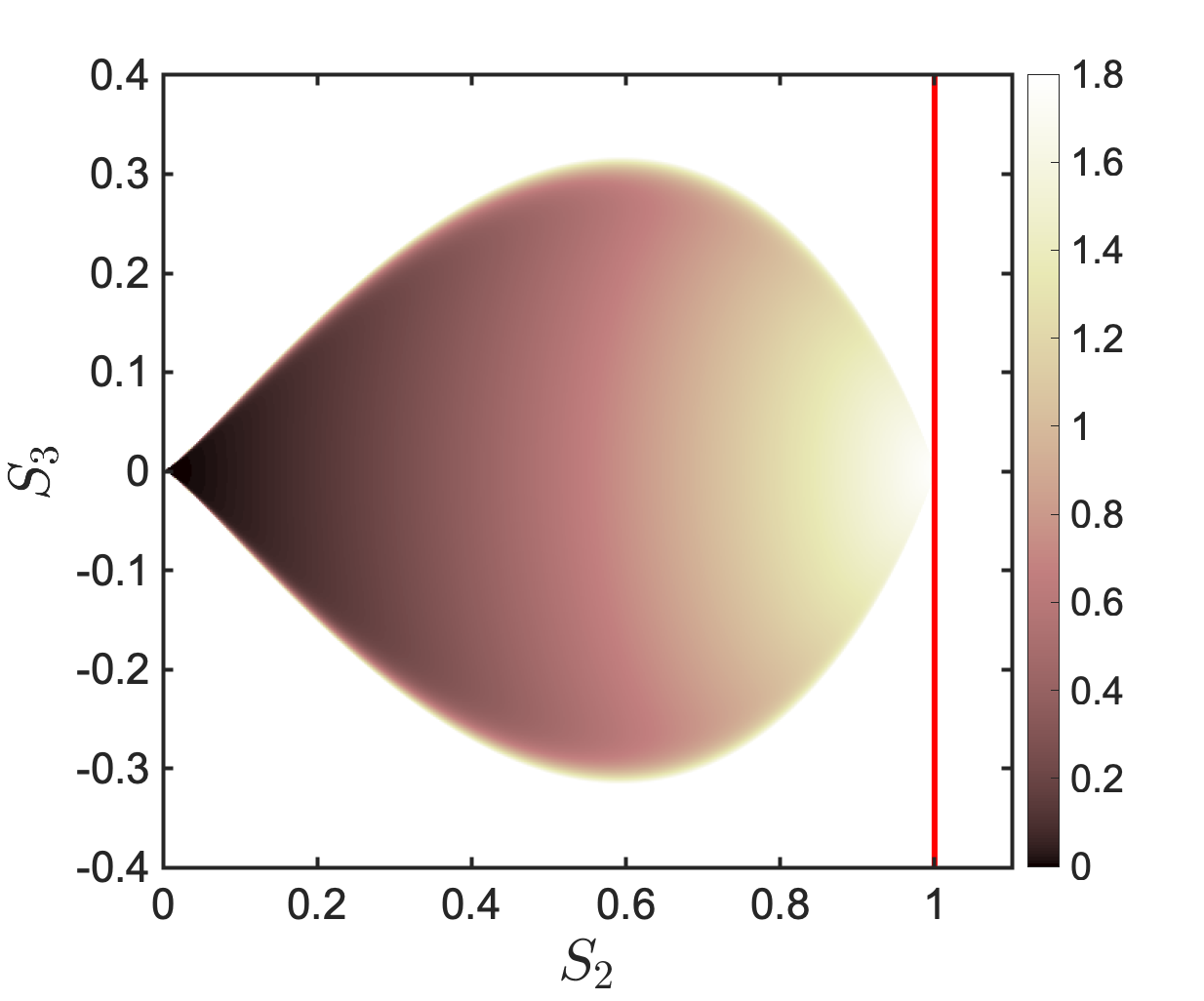}
\end{center}
  \caption{Contourplot of $S_4$ given by Eq.~(\ref{eqn:S4nc}) as a function of $S_2$ and $S_3$ for $\kappa=1$.  The summation is computed up to $S_3^{30}$.  The vertical red line corresponds to the
  location of the singularity at $S_2=S_2^{(c)}$.  The MATLAB~\cite{matlab} code to compute symbolically the terms of the closure and obtain numerically this figure is available at
  Ref.~\onlinecite{github}.}
  \label{fig:S4}
\end{figure}

{\em Remark: Scaling}. We notice that the functions $f_n$ satisfy
$$
f_n(\lambda^2S_2;\lambda^{2/3}\kappa)=\lambda^{4-6n} f_n(S_2;\kappa),
$$
for all $n\geq 0$. Therefore, we have a scaling relationship for $S_4$:
$$
S_4(\lambda^2 S_2, \lambda^3 S_3; \lambda^{2/3}\kappa)=\lambda^4 S_4(S_2,S_3;\kappa). 
$$
A contour plot of $S_4$ in the plane $(S_2,S_3)$ is represented in Fig.~\ref{fig:S4} for $\kappa=1$.
The equations of motion are given by Eqs.~(\ref{eqn:motion}) with
$$
R(\rho, P, q)=\rho^5\sum_{n=0}^\infty f_n\left(\frac{P}{\rho^3} \right) \left(\frac{2q}{\rho^4}\right)^{2n}.
$$
In particular one interesting feature is that the first order of the closure does not depend on $\rho$, i.e.,
$$
R(\rho,P,q=0)=\frac{9\kappa}{5} P^{5/3}. 
$$

{\em Remark: Relation between the kurtosis and the skewness}.  A scaling of kurtosis (related to $S_4$) with squared skewness (related to $S_3$) for plasma density fluctuations and sea-surface
temperature fluctuations was found in Refs.~\onlinecite{Labit07,Sura08,Krommes08,Sandberg09,Guszejnov13}.  Using the Hamiltonian closure, this relation is found as the first two terms of the closure,
i.e., $S_4=b + a S_3^2+O(S_3^4)$ where $a$ and $b$ are functions of $\rho$ and $P$.

\subsection{Casimir invariants}

A very interesting property of the noncanonical Poisson bracket~\eqref{eqn:PB} is that it possesses a number of Casimir invariants, i.e., observables $C$ such that $\{C,F\}=0$ for any other observable $F$.
First we are looking for Casimir invariants of the entropy type, i.e.,
$$
C=\int {\rm d}x \rho \Gamma(S_2,S_3).
$$
The function $\Gamma$ satisfies the following conditions:
\begin{equation}
\label{eqn:beta}
\beta_{ij}\frac{\partial^2 \Gamma}{\partial S_i \partial S_n}+\frac{\partial \gamma_{ij}}{\partial S_n}\frac{\partial\Gamma}{\partial S_i}=0,
\end{equation}
for all $j$, $k$ and $n$ in $(2, 3)$ (and where we assumed implicit summation over the repeated index $i$).  We assume that we have $K$ solutions, denoted $\Gamma_k$ for $k=2,\ldots,K$.  Using the
property $\beta=\gamma+\gamma^T$, we prove that the above-conditions are equivalent to
\begin{equation}
    \frac{\partial}{\partial S_n}\left(\frac{\partial \Gamma_k}{\partial S_i} \beta_{ij}\frac{\partial \Gamma_l}{\partial S_j}\right)=0, \label{eqn:betaconst}
\end{equation}
for all $n$, $k$ and $l$.

Using series expansions, we found two solutions to Eq.~\eqref{eqn:betaconst}:
\begin{eqnarray*}
&& C_2=\sum_{n=0}^\infty \int {\rm d}x \rho g_n(S_2) S_3^{2n},\\
&& C_3=\sum_{n=0}^\infty \int {\rm d}x \rho h_n(S_2) S_3^{2n+1},
\end{eqnarray*}
where the first elements in the series are:
\begin{eqnarray*}
&& g_0(S_2)=S_2^{1/3},\\
&& g_1(S_2)=-\frac{1}{3^3 S_2^{7/3} (\kappa - S_2^{1/3})},\\
&& g_2(S_2)=\frac{-4 \kappa + 3 S_2^{1/3}}{3^6 S_2^5 (\kappa - S_2^{1/3})^3},\\
&& h_0(S_2)= \frac{1}{3 S_2( \kappa - S_2^{1/3})}  ,\\
&& h_1(S_2)=  \frac{2}{3^4 S_2^{11/3}(\kappa - S_2^{1/3})^2},\\
&& h_2(S_2)= \frac{2(5\kappa -4S_2^{1/3})}{3^7 S_2^{19/3}( \kappa - S_2^{1/3})^4}.
\end{eqnarray*}
The functions $h_n$ and $g_n$ for $n\geq 1$ are determined from the recurrence relations:
\begin{eqnarray*}
&& g_{n+1}=-\frac{1}{9(n+1)(2n+1)S_2^{5/3}(\kappa - S_2^{1/3})}\left[(4n+1)g^\prime_n+\sum_{m=0}^n m(6n+4m+1) f_{n+1-m}g_m \right],\\
&& h_{n+1}=-\frac{1}{9(n+1)(2n+3)S_2^{5/3}(\kappa-S_2^{1/3})}\left[(4n+3)h^\prime_n+\sum_{m=0}^n (2m+3n+3)(2m+1)f_{n+1-m}h_m \right],
\end{eqnarray*}
which are both obtained from Eq.~\eqref{eqn:beta} with $j=3$ and $n=3$. 

These Casimir invariants allow us to define particularly relevant variables, referred to as normal variables, in which the Hamiltonian system is greatly simplified. 
We perform a local change of variables: $(S_2,S_3)\mapsto (\Gamma_2,\Gamma_3)$, where
\begin{eqnarray*}
&& \Gamma_2=\sum_{n=0}^\infty g_n(S_2) S_3^{2n},\\
&& \Gamma_3=\sum_{n=0}^\infty h_n(S_2) S_3^{2n+1}.
\end{eqnarray*} 
The bracket~\eqref{eqn:PB} becomes
\begin{eqnarray}
\{F,G\}&=&\int {\rm d}x \left[\fd G,u;\,\partial_x \fd F,\rho; - \fd F,u;\, \partial_x \fd G,\rho;\, -4\pi e \left(\fd G,u;\,\widetilde{\fd F,E;} - \fd F,u;\,\widetilde{\fd G,E;}\right) \right. 
\nonumber\\[4pt]
&& \qquad\left. -\frac{\partial_x \Gamma_i}{\rho}\left(\fd G,u;\,\fd F,\Gamma_i; - \fd F,u;\, \fd G,\Gamma_i;\right) + \frac1{\rho}\,\partial_x\left(\frac1{\rho}\,\fd 
F,\Gamma_i;\right)\tilde{\beta}_{ij}\fd G,\Gamma_j;\right], \label{eqn:PB2}
\end{eqnarray}
where $\tilde{\beta}$ is a symmetric matrix whose elements are 
$$
\tilde{\beta}_{kl}=\frac{\partial \Gamma_k}{\partial S_i}\beta_{ij}\frac{\partial \Gamma_l}{\partial S_j},
$$ 
with an implicit summation over repeated indices.  From Eq.~\eqref{eqn:betaconst}, we deduce that the matrix $\tilde{\beta}$ is constant.  As a consequence, the bracket~(\ref{eqn:PB2}) always
satisfies the Jacobi identity.  Therefore the existence of two Casimir invariants of the entropy type for the bracket~(\ref{eqn:PB}) is sufficient to ensure that it is a Poisson bracket.  Note that we
use the terminology Casimir invariant also for a bracket which is a priori not of the Poisson type.  Using the expressions for $S_3=0$, the matrix $\tilde{\beta}$ takes the very simple form
$$
\tilde{\beta}=\left(\begin{array}{cc}
0 & 1 \\ 1 & 0
\end{array} \right).
$$
In addition, the existence of two Casimir invariants of the entropy type ensures a third Casimir invariant:
$$
C_1=\int {\rm d}x \left(u-\frac{\rho}{2}\Gamma_k (\tilde{\beta}^{-1})_{kl} \Gamma_l  \right),
$$
which is equal to
$$
C_1=\int {\rm d}x \left(u-\rho\Gamma_2\Gamma_3  \right).
$$
Its expansion is given by
$$
C_1=\int {\rm d}x \left(u-\rho\sum_{n=0}^\infty k_n(S_2)S_3^{2n+1} \right),
$$
where
$$
k_{n}=\sum_{m=0}^n g_{n-m}h_m,
$$
for $n\geq 0$, and the first elements of the series are given by
\begin{eqnarray*}
&& k_0(S_2)= \frac{1}{S_2^{2/3}(\kappa - S_2^{1/3})} , \\
&& k_1(S_2)= \frac{1}{3^3 S_2^{10/3}(\kappa - S_2^{1/3})^2},\\ 
&& k_2(S_2)= \frac{4\kappa- 3 S_2^{1/3}}{3^6 S_2^{6}(\kappa - S_2^{1/3})^4}.
\end{eqnarray*}
We notice that three Casimir invariants similar to $C_1$, $C_2$ and $C_3$ (but of course, different) have been found for the Hamiltonian closure obtained using the dimensional analysis (see
Ref.~\onlinecite{Perin15}).

The advantage of working in the variables $\Gamma_i$ instead of the variables $S_i$ is that the closure functions $S_4$ and $S_5$ are no longer present in the Poisson bracket.  They are now in the
Hamiltonian through the change of variables $(S_2,S_3)\mapsto (\Gamma_2,\Gamma_3)$.  If we truncate the closure functions $S_4$ and $S_5$ --a natural step since these functions are given as series in
$S_3$-- the system remains Hamiltonian in the variables $\Gamma_i$ whereas if these truncations are performed in the bracket in the variables $S_i$, the system would likely loose the Hamiltonian
property.

\subsection{Parametric expression for the Hamiltonian closure}

There is another significant advantage to working with normal variables $\Gamma_i$: What is not fully satisfactory with the variables $S_i$ is that the closure is given as a relatively complex
expansion, and consequently we were not been able to check the Jacobi identity at all orders in the expansion.  The origin of this complication is due to the search for an
explicit closure function $S_4(S_2,S_3)$, not to the search of a Hamiltonian closure per se.  Here instead we are looking at a parametric expression of the closure, and we consider the normal
variables as parameters of the closure.  More precisely, we consider an arbitrary change of coordinates from some variables $\Gamma_i$ to variables $S_i$:
\begin{eqnarray*}
S_2=\overline{S_2}(\Gamma_2,\Gamma_3),\\
S_3=\overline{S_3}(\Gamma_2,\Gamma_3),
\end{eqnarray*}
and the closure functions are given by
\begin{eqnarray*}
S_4=\overline{S_4}(\Gamma_2,\Gamma_3),\\
S_5=\overline{S_5}(\Gamma_2,\Gamma_3). 
\end{eqnarray*}
We start with the bracket~\eqref{eqn:PB2} which is a Poisson bracket since the matrix $\tilde{\beta}$ is constant.  The question of finding Hamiltonian closures is reformulated as follows: What are
the functions $\overline{S_i}$ for which the bracket~\eqref{eqn:PB2} expressed in the variables $S_i$ is the original bracket~\eqref{eqn:PB}?  The answer is given by two sets of equations
\begin{eqnarray}
\frac{\partial \overline{S_i}}{\partial \Gamma_k}\widetilde{\beta}_{kl} \frac{\partial \overline{S_i}}{\partial \Gamma_l}=\beta_{ij}, \label{eqn:paramS}\\
\frac{\partial^2 \overline{S_i}}{\partial \Gamma_n \partial \Gamma_k}\widetilde{\beta}_{kl} \frac{\partial \overline{S_i}}{\partial \Gamma_l} 
= \frac{\partial \gamma_{ij}}{\partial \Gamma_n}, \label{eqn:constS}
\end{eqnarray}
for all $i$, $j$ and $n$. The first set of equations~\eqref{eqn:paramS} defines parametrically the functions $\overline{S_3}$, $\overline{S_4}$ and $\overline{S_5}$:
\begin{eqnarray*}
&& \overline{S_3}=\frac{1}{2}\frac{\partial \overline{S_2}}{\partial \Gamma_2}\frac{\partial \overline{S_2}}{\partial \Gamma_3},\\
&& \overline{S_4}=\frac{9}{5} \overline{S_2}^2+\frac{1}{5}\frac{\partial \overline{S_2}}{\partial \Gamma_2}\frac{\partial \overline{S_3}}{\partial \Gamma_3}
+ \frac{1}{5}\frac{\partial \overline{S_2}}{\partial \Gamma_3}\frac{\partial \overline{S_3}}{\partial \Gamma_2},\\
&& \overline{S_5}=4\overline{S_2}\ \overline{S_3}+ \frac{1}{3}\frac{\partial \overline{S_3}}{\partial \Gamma_2}\frac{\partial \overline{S_3}}{\partial \Gamma_3}.
\end{eqnarray*}
Once the function $\overline{S_2}$ is specified, all of the other functions $\overline{S_i}$ are uniquely determined by the above equations.  By inverting the equations $\Gamma_i=\Gamma_i(S_2,S_3)$ or
by solving one of the constraints~\eqref{eqn:constS}, we obtain the following expression for $\overline{S_2}(\Gamma_2,\Gamma_3)$:
\begin{equation}
    \overline{S_2}(\Gamma_2,\Gamma_3)=\Gamma_2^3+\Gamma_2(\kappa-\Gamma_2)\Gamma_3^2\label{eqn:S2p}.
\end{equation}
Inserting this expression in the parametric equations for $S_3$, $S_4$ and $S_5$ leads to the following expressions:
\begin{subequations}
\label{eqn:S4param}
\begin{align}
& \overline{S_3}(\Gamma_2,\Gamma_3)=\Gamma_2 \Gamma_3(\kappa-\Gamma_2)\left(3\Gamma_2^2+(\kappa-2\Gamma_2)\Gamma_3^2\right),\label{eqn:S3p}\\
& \overline{S_4}(\Gamma_2,\Gamma_3)=\frac{9\kappa}{5} \Gamma_2^5+6\Gamma_2^3(\kappa-\Gamma_2)^2\Gamma_3^2\nonumber \\
& \qquad \qquad  +\Gamma_2(\kappa-\Gamma_2)\left(\kappa^2-3\Gamma_2(\kappa-\Gamma_2)\right)\Gamma_3^4,\label{eqn:S4p}\\
& \overline{S_5}(\Gamma_2,\Gamma_3)=9\kappa\Gamma_2^5(\kappa-\Gamma_2)\Gamma_3+10\Gamma_2^3(\kappa-\Gamma_2)^3\Gamma_3^3\nonumber\\
& \qquad \qquad +\Gamma_2(\kappa-\Gamma_2)(\kappa-2\Gamma_2)(\kappa^2-2\kappa\Gamma_2+2\Gamma_2^2)\Gamma_3^5.\label{eqn:S5p}
\end{align}
\end{subequations}
We notice that the closure is no longer given as an infinite series.  In particular, the functions $\overline{S_n}$ for $n=2,3,4,5$ are polynomials in the two variables $\Gamma_2$ and $\Gamma_3$, and
the degree in $\Gamma_3$ is $n$ and the degree in $\Gamma_2$ is $n+1$.  Using Mathematica~\cite{mathematica}, we have checked that the constraints~\eqref{eqn:constS} are all satisfied.  The code is
available at Ref.~\onlinecite{github}.  The series expansion of the explicit closure $S_4=S_4(S_2,S_3)$ given in Eqs.~\eqref{eqn:S4nc} is obtained by inverting Eqs.~\eqref{eqn:S2p} and
\eqref{eqn:S3p}, and inserting them in Eq.~\eqref{eqn:S4p}. As a consequence, this proves the Jacobi identity for the explicit closure $S_4=S_4(S_2,S_3)$. 

For $S_2$ to be positive, a necessary and sufficient condition is that $\kappa>\Gamma_2>0$ or if $\Gamma_2>\kappa$, $\Gamma_3^2< \Gamma_2^2/(\Gamma_2-\kappa)$.  This means that $S_2$ can take
arbitrarily large values, provided that $S_3$ is not too large.  We notice that the point $(S_2=\kappa, S_3=0)$ in Fig.~\ref{fig:S4} is obtained for $\Gamma_2=\kappa$ regardless of the value of
$\Gamma_3$.

\begin{figure}
\begin{center}
\includegraphics[width=0.7\linewidth]{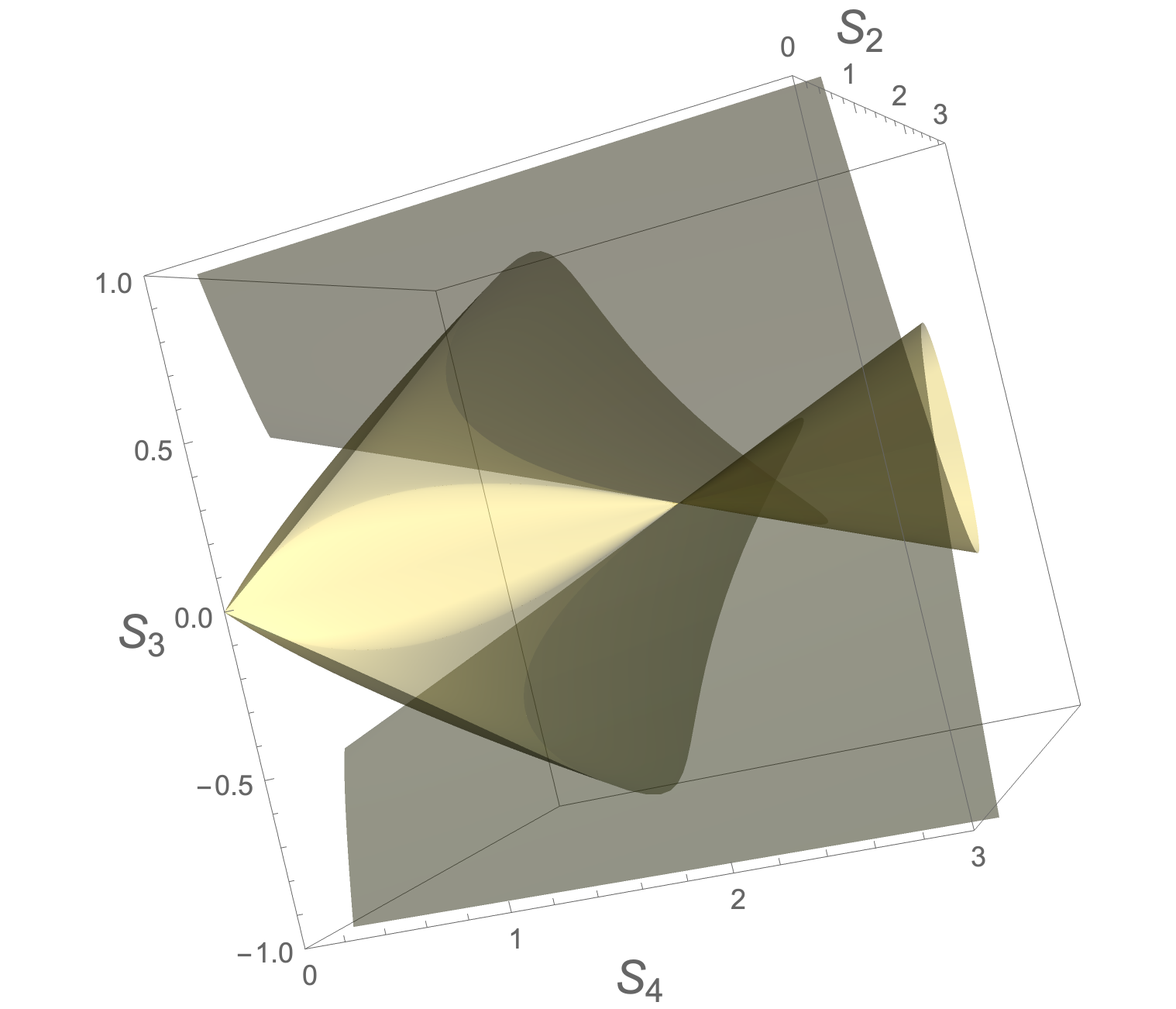}
\end{center}
  \caption{Parametric representation of $S_4$ given by Eqs.~(\ref{eqn:S4param}) as a function of $S_2$ and $S_3$ for $\kappa=1$. The Mathematica code is available at Ref.~\onlinecite{github}}
  \label{fig:S4p}
\end{figure}

In Fig.~\ref{fig:S4p}, we have represented the closure function $S_4$ given parametrically by Eqs.~\eqref{eqn:S4param} for a selected range of parameters $(\Gamma_2,\Gamma_3)$.  The surface gets more
complicated, with more branches, as the range of $(\Gamma_2,\Gamma_3)$ is extended (see the Mathematica code available at Ref.~ \onlinecite{github}).
We notice that there is a central brighter patch where there is a single value of $S_4$ for a given $(S_2,S_3)$.  It corresponds to the explicit closure $S_4=S_4(S_2,S_3)$ as depicted in
Fig.~\ref{fig:S4}.

\subsection{Equations of motion}

The Poisson bracket~\eqref{eqn:PB2} becomes 
\begin{eqnarray*}
\{F,G\}&=&\int {\rm d}x \left[\fd G,u;\, \partial_x \fd F,\rho;\, -\fd F,u;\, \partial_x \fd G,\rho;\, -4\pi e\left(\fd G,u;\,\widetilde{\fd F,E;} - \fd F,u;\,\widetilde{\fd G,E;\,}\right) \right. \nonumber\\
&& \qquad\left.{} - \frac{\partial_x \Gamma_i}{\rho}\left(\fd G,u;\,\fd F,\Gamma_i; -\fd F,u;\,\fd G,\Gamma_i;\right)
+ \frac1{\rho}\,\fd G,\Gamma_2;\,\partial_x\left(\frac1{\rho}\,\fd F,\Gamma_3;\right) - \frac1{\rho}\,\fd F,\Gamma_2;\,\partial_x\left(\frac1{\rho}\,\fd G,\Gamma_3;\right) \right],
\end{eqnarray*}
and the Hamiltonian is
$$
H[\rho,u,\Gamma_2,\Gamma_3,E]=\frac{1}{2}\int {\rm d}x \left[ \rho u^2 +\rho^3 \overline{S_2}(\Gamma_2,\Gamma_3)+\frac{E^2}{4\pi} \right],
$$
where $\overline{S_2}$ is given by Eq.~\eqref{eqn:S2p}.
The equations of motion are given by $\dot{F}=\{F,H\}$:
\begin{subequations}
\label{eqn:motionG}
\begin{eqnarray}
&&  {\partial_t} \rho=-{\partial_x} (\rho u),\label{eqn:motionGa}\\
&&  {\partial_t} u=-u {\partial_x} u-\frac{1}{\rho}\partial_x \left(\rho^3 \overline{S_2} \right) +e\widetilde{E}\label{eqn:motionGb}\\
&&  {\partial_t}\Gamma_2= -u {\partial_x}\Gamma_2-\frac{1}{2\rho}\partial_x\left(\rho^2\frac{\partial \overline{S_2}}{\partial \Gamma_3} \right),\label{eqn:motionGc}\\
&&  {\partial_t}\Gamma_3= -u {\partial_x}\Gamma_3-\frac{1}{2\rho}\partial_x\left(\rho^2\frac{\partial \overline{S_2}}{\partial \Gamma_2} \right),\label{eqn:motionGd}\\
&&  \partial_t E=-4\pi e \widetilde{\rho u}.\label{eqn:motionGe}
\end{eqnarray}
\end{subequations}

{\em Remark 1:} In the case of an external time-dependent electric field $E_0(x,t)$, the closure is identical.  First we need to autonomize the bracket.  For the Vlasov--Poisson equation, the variables
are the fields $f(x,v,t)$ and $E_1(x,t)$, together with $t$ and $K$ ($K$ being the canonically conjugate variable to time $t$), such that the total electric field is $E=E_0+E_1$.  The Hamiltonian is
$$
H[f,E_1,t,K]=\int {\rm d}x {\rm d}v f\frac{v^2}{2}+\int {\rm d}x \frac{E_1^2+2E_1 E_0}{8\pi}+ K,
$$
and the Poisson bracket
\begin{eqnarray*}
\{F,G\}&=&\int f\left[\partial_x \fd F,f;\,\partial_v\fd G,f; - \partial_v \fd F,f;\,\partial_x \fd G,f;\, 
-4\pi e \left(\widetilde{\fd F,E_1;\,}\partial_v \fd G,f; - \partial_v \fd F,f;\,\widetilde{\fd G,E_1;} \right) \right]{\rm d}x {\rm d}v\\
&& {}+F_tG_K-F_K G_t.
\end{eqnarray*}
For the reduced fluid equations, the Hamiltonian becomes
$$
H[\rho, u, \Gamma_2,\Gamma_3, E_1,t,K]= \frac{1}{2}\int {\rm d}x \left[\rho u^2+\rho^3 \overline{S_2} +\frac{E_1^2+2E_1 E_0}{4\pi}\right]+K,
$$
and the Poisson bracket
\begin{eqnarray*}
\{F,G\}&=&\int {\rm d}x \left[\fd G,u;\, \partial_x \fd F,\rho;\, - \fd F,u;\, \partial_x \fd G,\rho;\, -4\pi e\left(\fd G,u;\, \widetilde{\fd F,E_1;}-\fd F,u;\, \widetilde{\fd G,E_1;}\right) \right. \nonumber \\
&& \qquad\left. {}-\frac{\partial_x \Gamma_i}{\rho}\left(\fd G,u;\,\fd F,\Gamma_i; -\fd F,u;\,\fd G,\Gamma_i;\right)
+  \frac1{\rho}\,\fd G,\Gamma_2;\,\partial_x \left(\frac1{\rho}\,\fd F,\Gamma_3;\right) - \frac1{\rho}\,\fd F,\Gamma_2;\,\partial_x\left(\frac1{\rho}\,\fd G,\Gamma_3;\right)\right]\\
&&{} + F_t G_K - F_K G_t.
\end{eqnarray*}
The equations of motion consists in changing $E$ by $E_0+E_1$ in the Vlasov equation and in the momentum equation, and replacing $E$ by $E_1$ in the Amp\`ere equation. 

{\em Remark 2:} By rescaling the parameters $\Gamma_2$ and $\Gamma_3$, and by rescaling the density $\rho$ in the following way
\begin{eqnarray*}
&& \Gamma_2=\kappa\,\Gamma^{(\rm r)}_2,\\
&& \Gamma_3=\sqrt{\kappa}\,\Gamma^{(\rm r)}_3,\\
&& \rho = \kappa^{-3/2}\rho^{(\rm r)},
\end{eqnarray*}
the equations of motion~\eqref{eqn:motionGa}-\eqref{eqn:motionGd} are not longer explicitly depending on $\kappa$.  The parameter $\kappa$ appears only in Amp\`ere's equation or equivalently in Gauss'
law.  This means that the parameter of the closure $\kappa$ can be viewed as the coupling parameter between the fluid part and the electrostatic part.  The parameter $\kappa$ can also be removed
completely from the equations of motion by rescaling the charge and the electric field as
\begin{eqnarray*}
&& e = \kappa^{3/4} e^{(r)},\\
&& E = \kappa^{-3/4} E^{(r)}.
\end{eqnarray*}
As a consequence, the one-parameter family of Hamiltonian closures can be seen as a unique Hamiltonian model, and the parameter $\kappa$ is now in the initial condition.  

\subsection{Stability of the symmetric and homogeneous equilibria}

We have found a one-parameter family of closures which fulfill the first requirement, namely, the resulting models are Hamiltonian.  The second requirement is the stability of the equilibria $q_0=0$.
The linearized equations of motion reduce to Eq.~(\ref{eqn:lin}) with
$$
A=\left( 
\begin{array}{ccccc}
0 & -i k \rho_0 & 0 & 0 & 0\\
0 & 0 & -i k \rho_0^{-1} & 0 & e/m \\
0 & -3i kP_0 & 0 & -2 i k & 0\\
0 & 0 & -3ik(\kappa P_0^{2/3}-\rho_0^{-1}P_0)/2 & 0 & 0\\
0 & -4\pi e\rho_0 & 0 & 0 & 0
\end{array} \right).
$$
From the dispersion relation, we define
$$
{\omega}_{0}^2=\omega_p^2+3\kappa P_0^{2/3}k^2,
$$
where $\omega_p=\sqrt{4\pi e^2 \rho_0/m}$ is the plasma frequency.
The eigenvalues of $A$ are all purely imaginary if
$$
{\omega}_{0}^2>\omega_{BG}^2,
$$
where $\omega_{BG}(k)$ is the Bohm-Gross dispersion relation given by
$$
\omega_{BG}^2=\omega_p^2+3\frac{P_0}{\rho_0}k^2.
$$
The non-zero eigenvalues of $A$ are
$$
i \omega=\pm \frac{i }{\sqrt{2}} \left({\omega}_{0}^2\pm\sqrt{{\omega}_{0}^4-4\omega_p^2 ({\omega}_{0}^2-\omega_{BG}^2)} \right)^{1/2}.
$$
Therefore the homogeneous equilibria are stable for $\omega_0^2>\omega_{BG}^2$, which is equivalent to requiring that $S_2<S_2^{(c)}$ or $\Gamma_2<\kappa$.  In terms of the parameters of the
equilibrium, this means that the pressure $P_0$ is such that $P_0^{1/3}/\rho_0 <\kappa$.  A crucial factor is that the closure $R(\rho,P,q=0)$ does not depend on $\rho$, and in this case, the
necessary and sufficient condition for stability is
$$
\omega_p^2+k^2\frac{\partial R}{\partial P}>\omega_{BG}^2.
$$
We recall that 
$$
R(\rho,P,0)=\rho^5 S_4\left(\frac{P}{\rho^3} ,0\right).
$$
The fractional exponent $5/3$ in the closure comes from the requirement that $R$ does not depend on $\rho$, ensuring the stability of the equilibria.  More general cases for stability would be that at
$q=0$
\begin{eqnarray*}
&& \frac{\partial R}{\partial P}> \frac{3P}{\rho},\\
&& \frac{\partial R}{\partial \rho }\leq 0,
\end{eqnarray*}
for all $\rho>0$ and $P>0$.  However, these conditions do not ensure that the resulting model is Hamiltonian.  As expected, the requirement that the model is Hamiltonian is more stringent than
requiring that homogeneous equilibria are stable.

\section{Numerical applications}

The objective of this section is not to offer a detailed comparison between the numerical implementation of the Hamiltonian fluid model and the one of the parent kinetic model.  The objective is more
modest since we limit ourselves to a couple of illustrations of the Hamiltonian fluid model, demonstrating the feasibility and practicality of the fluid model, which could trigger further questions of
a more practical nature than the ones we consider in what follows.  We consider two applications, one where the fluid model leads to stable plasma oscillations and the other one where it is unstable.
In all the simulations, we consider a domain $x\in [-L_x,L_x]$ and $v\in [-L_v,L_v]$ with $L_v=10$.

\subsection{Plasma oscillations}

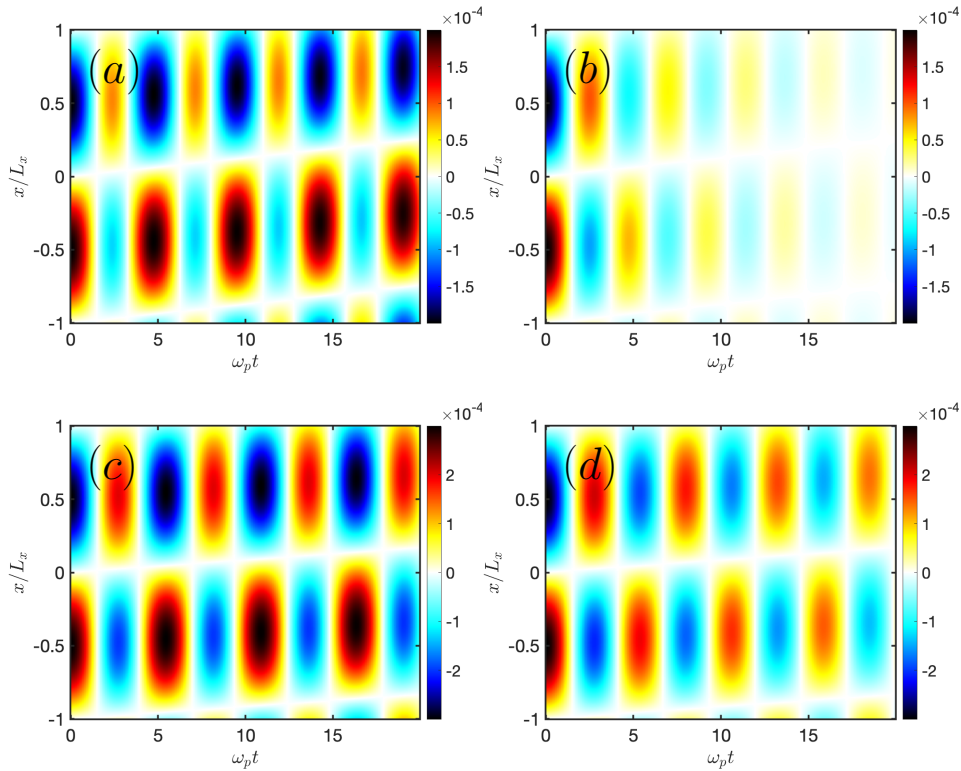
\begin{figure}
\begin{picture}(300,300)
\put(0,150){\includegraphics[width=0.4\linewidth]{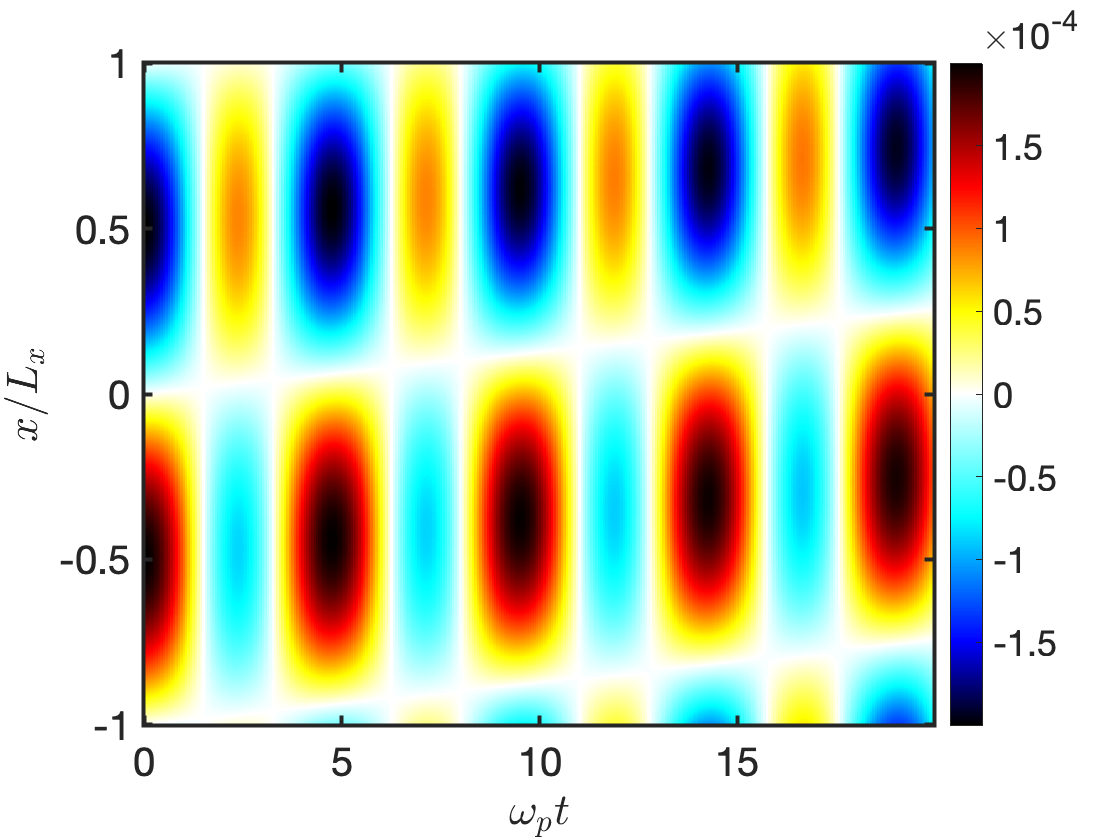}}
\put(180,150){\includegraphics[width=0.4\linewidth]{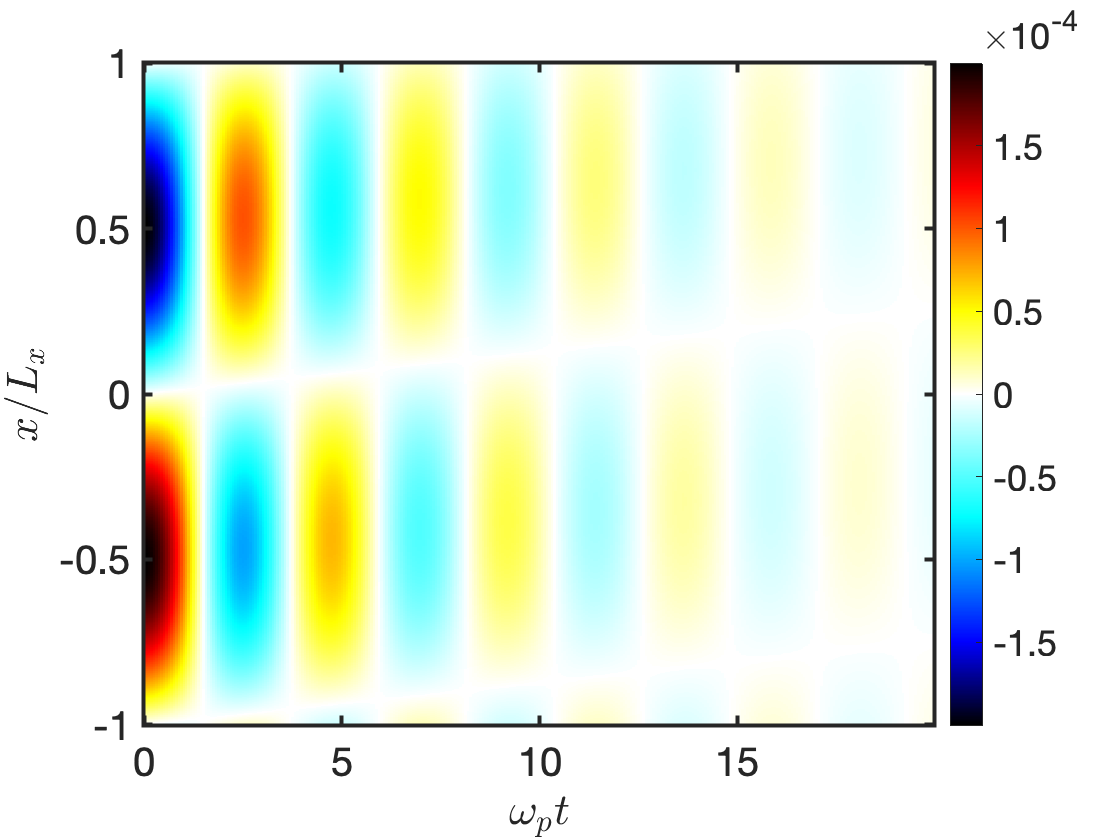}}
\put(0,0){\includegraphics[width=0.4\linewidth]{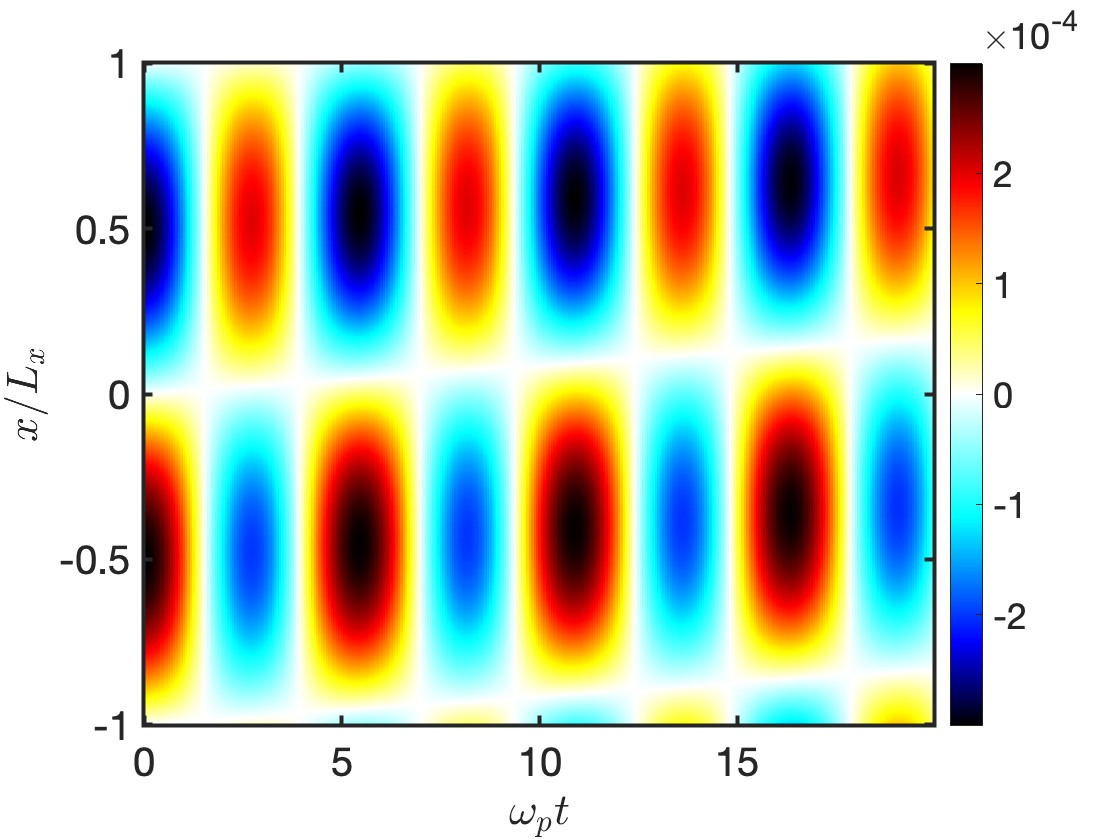}}
\put(180,0){\includegraphics[width=0.4\linewidth]{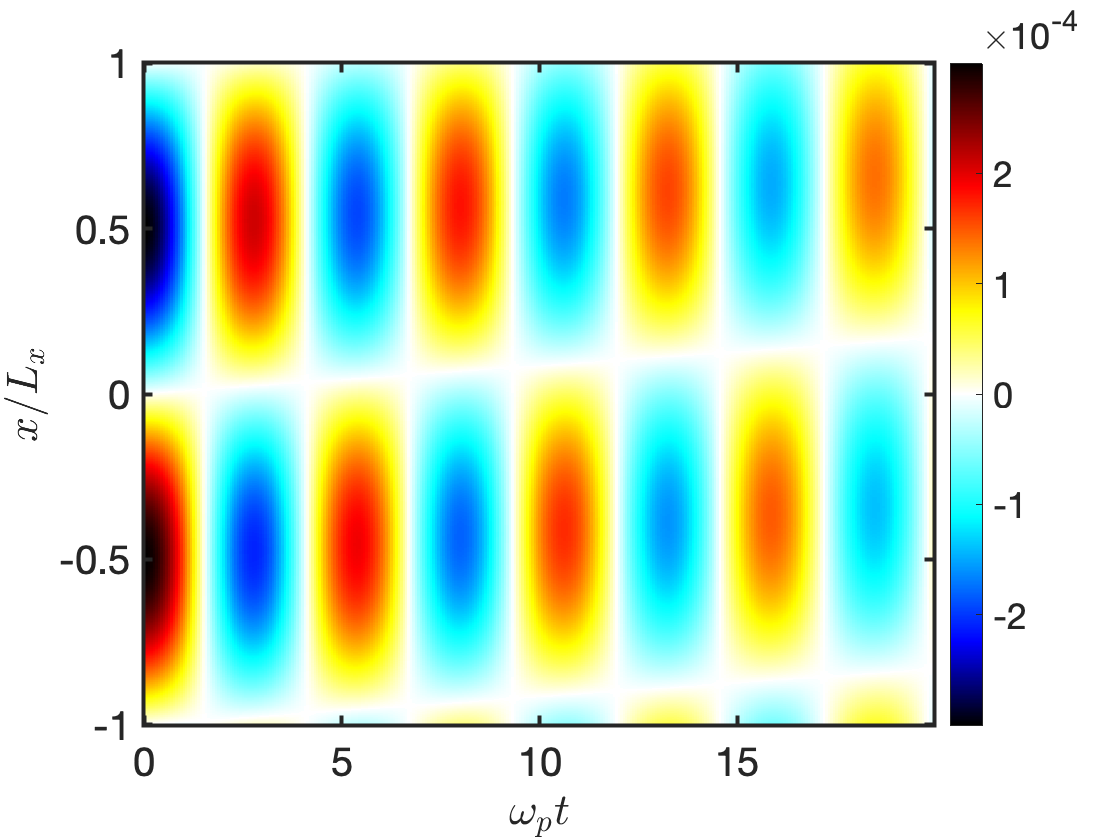}}
\put(30,260){\Large $(a)$}
\put(210,260){\Large $(b)$}
\put(30,110){\Large $(c)$}
\put(210,110){\Large $(d)$}
\end{picture}
  \caption{Contour plot of $E(x,t)$: Panel $(a)$: Hamiltonian fluid model with $\kappa=1$ and $L_x/\lambda_D=2\pi$. Panel $(b)$: one-dimensional Vlasov--Poisson equation with $L_x/\lambda_D=2\pi$. 
  Panel $(c)$: Hamiltonian fluid model with $\kappa=1$ and $L_x/\lambda_D=3\pi$.
  Panel $(d)$: one-dimensional Vlasov--Poisson equation with $L_x/\lambda_D=3\pi$.}
  \label{fig:N1}
\end{figure}

We consider the following initial distribution function, built from a skew-normal distribution,
$$
f(x,v,0)=\frac{1}{\sqrt{2\pi}}\,(1-A\cos k\,x) \left[1+{\rm erf}\left( \frac{\alpha v}{\sqrt{2}}\right) \right]{\rm e}^{-v^2/2} ,
$$
with $A=10^{-4}$, $k=\lambda_D/2$ and $\alpha=0.1$ (where $\lambda_D$ is the Debye length). Here the velocities are in units of the thermal velocity $v_{\rm th}=\sqrt{k_B T}$.
Given that the equilibrium has some initial fluid velocity, the Bohm-Gross dispersion relation is becomes
$$
\omega_{BG}= \pm \omega_p\left[1 \pm u\frac{k}{\omega_p}+\frac{3}{2}\rho^2{S}_2 \frac{k^2}{\omega_p^2} \pm 2\rho^3 {S}_3\frac{k^3}{\omega_p^3} +O\left(\frac{k^4}{\omega_p^4} \right)\right].
$$
This is the same dispersion relation given by the fluid and the kinetic models. For the skew-normal equilibrium, 
\begin{eqnarray*}
&& \rho=1,\\
&& u=\alpha \sqrt{\frac{2}{\pi(1+\alpha^2)}},\\
&& S_2= 1-\frac{2\alpha^2}{\pi (1+\alpha^2)},\\
&& S_3=\frac{\alpha^3}{(1+\alpha^2)^{3/2}}\sqrt{\frac{2}{\pi}}\left(\frac{4}{\pi}-1\right).
\end{eqnarray*}
We consider the fluid model with $\kappa=1$.  Given the initial values of $S_2$ and $S_3$, we compute the initial values for $\Gamma_2$ and $\Gamma_3$.  We represent the values of $E(x,t)$ in
Fig.~\ref{fig:N1} obtained with the fluid and the kinetic model.
We notice some qualitative similarities between the kinetic and the fluid model, such as plasma oscillations.  However, as expected, the fluid model does not capture the damping of the field (clearly
visible for $L_x/\lambda_D=2\pi$), which is a purely kinetic effect.  For larger values of $L_x$, i.e., $L_x/\lambda_D=3\pi$ the damping is reduced as expected, and the agreement between the kinetic
and the fluid simulations is improved.

\subsection{Two-stream instability}

Next, we consider the two-stream instability with the initial distribution
$$
f(x,v,0)=(1-A\cos kx)\frac{v^2 {\rm e}^{-v^2/2v_0^2}}{\sqrt{2\pi}\,v_0^3}.
$$
For this distribution, $v_\mathrm{th}= \sqrt3\,v_0$.  To simplify comparison with the existing literature, we take $\lambdab = \lambda_D/\sqrt3$ and $v_0$ to be our length and velocity scales,
respectively.  We set $A=10^{-6}$ and $k\,\lambdab=1/2$.  From the previous section, we know that the Hamiltonian fluid model leads to an instability if $\kappa<3^{1/3}\approx 1.44$ (since $\rho_0=1$
and $P_0=3$).  Here we consider the fluid model with $\kappa=1.30834$.

\begin{figure}
\includegraphics{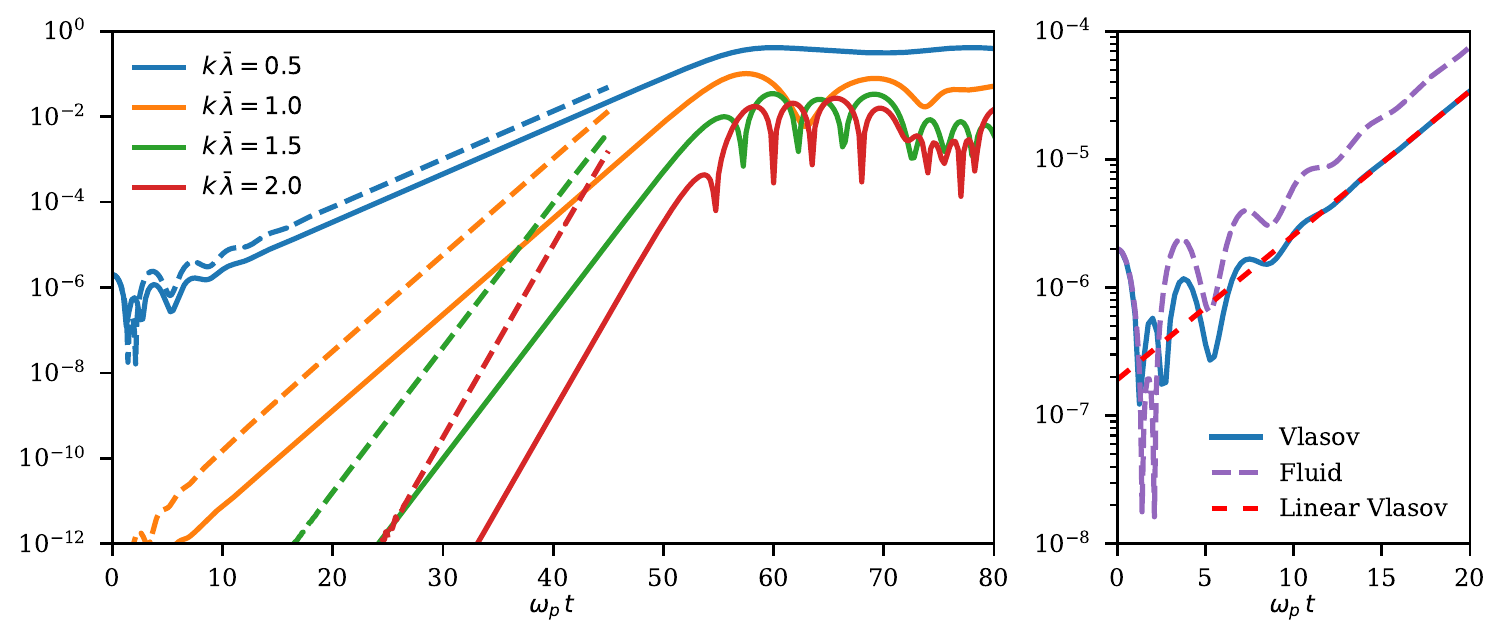}
  \caption{The left panel shows magnitude of the Fourier modes $E_k(t)$ of the electric field $E(x,t)$ as functions of time for $k\,\lambdab=1/2$ (blue curves), $k\,\lambdab=1$ (orange curves),
  $k\,\lambdab=3/2$ (green curves) and $k\,\lambdab=2$ (read curves).  The continuous curves are for the kinetic model, and the dashed curves are for the Hamiltonian fluid model with
  $\kappa=1.30834$.  The right panel shows the amplitudes for $k\,\lambdab=1/2$ for Hamiltonian fluid model (dashed violet), kinetic (blue) and unstable mode from linear kinetic theory (red).}
  \label{fig:N2}
\end{figure}

In Fig.~\ref{fig:N2}, we compare the growth of the first four Fourier modes of the electric field, i.e., with $k\,\lambdab=1/2$ (fundamental), $k\,\lambdab=1$, $k\,\lambdab=3/2$ and $k\lambdab=2$ for
$L_x/\lambda_D=2\pi$.
As expected, both models, fluid and kinetic, display the instability, i.e., the growth of the electric field with time.  The numerical algorithm for the fluid model fails at $\omega_p t\approx
47$, at which time particle trapping becomes predominant in the kinetic model.

The parameter $\kappa$ has been chosen such that the slope of the linear part of the first mode $k\,\lambdab=1/2$ obtained with the fluid model matches the one obtained with the linear kinetic model,
i.e., a growth rate of $0.25924553\,\omega_p$ (which has been corrected for the effects of the spatial grid).  We notice that both models display some similar features, such as the oscillations at the
beginning.  Also, the slope of the higher-order modes corresponds rather well, despite the fact that these modes are higher in amplitude for the fluid model.

The main discrepancy between both models occur when the amplitude of the field saturates, which is when the kinetic effects are predominant, and these cannot be described by the fluid model.  In
addition, all wavenumbers are unstable in the Hamiltonian fluid model while only the fundamental mode is unstable in the kinetic model (the higher harmonics are driven by the fundamental mode through
nonlinear couplings).  For both models, the initial electric field has the same initial amplitude.  Nonetheless, the amplitude of the fundamental mode is slightly larger in the fluid model compared to
the kinetic model (cf.  the blue curves on the left panel of Fig.~\ref{fig:N2}).  This is due to differences in how the initial condition projects onto the system modes in the two models.  In both
cases, a linear analysis produces mode amplitudes that are in excellent agreement with the numerical results.

\section*{Conclusions}
We have exhibited a one-parameter family of Hamiltonian fluid models with the first four fluid moments -- fluid density, fluid velocity, pressure and heat flux -- as a result of the reduction of the
one-dimensional Vlasov--Poisson equation.  The closure involves an equation for the kurtosis in velocity of the distribution function.  In the course of the reduction to a Hamiltonian fluid model, we
have identified some normal variables in which the closure expressed parametrically is found to be polynomial in the normal variables.  Each reduced Hamiltonian fluid model possesses three Casimir
invariants, two of the entropy type and one generalized velocity.  We have shown that some of these models ensures the stability of symmetric homogeneous equilibria, depending on the parameter of the
closure and the initial conditions.

\section*{Acknowledgments}
CC acknowledges useful discussions with J. F\'ejoz and BAS acknowledges useful discussions with Frank M. Lee.  This material is based upon work supported by the National Science Foundation under Grant
No.~DMS-1440140 while the authors were in residence at the Mathematical Sciences Research Institute in Berkeley, California, during the Fall 2018 semester.  This work has been carried out within the
framework of the French Federation for Magnetic Fusion Studies (FR-FCM).  BAS was supported in part by the National Science Foundation under Contract No.~PHY-1535678.

\section*{Data availability}
Data sharing is not applicable to this article as no new data were created or analyzed in this study.


%

\end{document}